\documentclass[useAMS,usegraphicx,usenatbib]{mn2e}
\usepackage[a4paper,centering, totalwidth=520pt, totalheight=700pt]{geometry}
\usepackage{graphicx}
\usepackage{aas_macros}     
\usepackage{amsmath}
\usepackage{amssymb}
\usepackage{fixltx2e}[1999/12/01]
\usepackage{bm}

\usepackage[hyperindex]{hyperref}
\hypersetup{
breaklinks = {true},
colorlinks = {false},
linkcolor={black},
pdfpagemode = {None}, 
pdfborder = {0 0 1},
pdftitle = {Noiseless Gravitational Lensing Simulations},
pdfsubject = {Noiseless Gravitational Lensing Simulations},
pdfauthor = {Raul E. Angulo, Ruizhu Chen, Stefan Hilbert, Tom Abel},
pdfkeywords = {gravitational lensing: strong -- gravitational lensing: weak -- cosmology: theory -- dark matter  -- large-scale structure of the Universe -- methods: numerical}
}

\newcommand{\Msun}{ \ensuremath{\mathrm{M}_{\odot}} }
\newcommand{\Mass}{ \ensuremath{ h^{-1} \Msun }}
\newcommand{\Mpc}{ \ensuremath{h^{-1} {\rm Mpc}} }

\newcommand{\hMpc}{ \ensuremath{ h\,{\rm Mpc^{-1}}} }
\newcommand{\kpc}{ \ensuremath{ h^{-1} {\rm kpc}} }
\newcommand{\eV}{ \ensuremath{\mathrm{eV}}}

\newcommand{\cic}{{CiC}}
\newcommand{\sph}{{Sph}}
\newcommand{\tcm}{{TCM}}
\newcommand{\rtcm}{{Recursive-TCM}}

\newcommand{\vect}[1]{\boldsymbol{#1}}

\newcommand{\ii}{\mathrm{i}}
\newcommand{\diff}[1]{\mathrm{d}#1}
\newcommand{\clight}{\ensuremath{\mathrm{c}}}
\newcommand{\GNewton}{\ensuremath{\mathrm{G}}}

\newcommand{\vtheta}{\ensuremath{\vect{\theta}}}
\newcommand{\vell}{\ensuremath{\vect{\ell}}}

\newcommand{\pdf}{\mathrm{pdf}}
\newcommand{\Sec}[1]{Section~\ref{#1}}

\usepackage[usenames,dvipsnames]{color}

\title[Noiseless Lensing Simulations]{Noiseless Gravitational Lensing Simulations}

\author[R. Angulo, R. Chen, S. Hilbert, T. Abel]{
    Raul E. Angulo\thanks{Email: reangulo@cefca.es}$^{1,2}$,
    Ruizhu Chen$^{2}$,
    Stefan Hilbert$^{3}$\&
    Tom Abel$^{2,4}$ \\
  $^{1}$Centro de Estudios de F\'isica del Cosmos de Arag\'on,  Plaza San Juan 1,  Planta-2, 44001, Teruel, Spain.\\
  $^{2}$Kavli Institute for Particle Astrophysics and Cosmology, \\
  \quad Stanford University, SLAC National Accelerator Laboratory,  Menlo Park, CA 94025, USA.\\
   $^3$ Max-Planck-Institute for Astrophysics, Karl-Schwarzschild-Str. 1, 85740 Garching, Germany. \\
   $^{4}$Institut Lagrange de Paris, Institut d'Astrophysique de Paris, 98 bis boulevard Arago - 75014 Paris, France. \\
    }

\begin{document}
\maketitle

\date{\today}
\pagerange{\pageref{firstpage}--\pageref{lastpage}} \pubyear{2013}
\label{firstpage}

\begin{abstract}
The microphysical properties of the DM particle can, in principle, be constrained
by the properties and abundance of substructures in DM halos, as measured
through strong gravitational lensing. Unfortunately, there is a lack of accurate
theoretical predictions for the lensing signal of substructures, mainly because
of the discreteness noise inherent to $N$-body simulations. Here we present \rtcm{},
a method that is able to provide lensing
predictions with an arbitrarily low discreteness noise, without any free
parameters or smoothing scale. This solution is based on a novel way of
interpreting the results of $N$-body simulations, where particles simply trace
the evolution and distortion of Lagrangian phase-space volume elements. We
discuss the advantages of this method over the widely used cloud-in-cells and
adaptive-kernel smoothing density estimators. Applying the new method to a
cluster-sized DM halo simulated in warm and cold DM scenarios, we show how the
expected differences in their substructure population translate into differences
in the convergence and magnification maps. We anticipate that our method will
provide the high-precision theoretical predictions required to interpret and
fully exploit strong gravitational lensing observations.

\end{abstract}

\section{Introduction}
\label{s:introduction}

Gravitational Lensing  has become a powerful and robust technique to explore
the ``dark side'' of our Universe \citep[see][for a recent
review]{Bartelmann:2010}. In the near future, it is expected to probe the
accelerated cosmic expansion, and to constrain the properties of the dark
matter (DM) particle.

In the weak regime, lensing by the large-scale structure of the Universe causes
small distortions in the apparent shape of galaxies and in the temperature
fluctuations maps at the last scattering surface. This effect can be detected
statistically by future wide-field surveys (e.g
DES\footnote{\url{http://www.darkenergysurvey.org/}},
J-PAS\footnote{\url{http://j-pas.org/}},
Euclid\footnote{\url{http://sci.esa.int/euclid/}},
LSST\footnote{\url{http://www.lsst.org/}}), and by Cosmic Microwave
Background (CMB) experiments (Planck\footnote{\url{http://www.esa.int/Our\_Activities/Space\_Science/Planck/}},
SPT\footnote{\url{http://pole.uchicago.edu/}},
ACT\footnote{\url{http://www.princeton.edu/act/}}).
From correlations in the
distortions, one can infer the amplitude, shape, and redshift evolution of the
matter power spectrum -- quantities sensitive to the initial density
perturbations, the law of gravity and the cosmic expansion. Therefore,
gravitational lensing measurements are expected to contribute significantly to
our understanding of the Dark Energy and the physics of the early Universe
\citep[e.g.][]{Huterer2010,MarianEtal2011,OguriTakada2011,Hilbert2012,Giannantonio2012}.

In the strong regime, efficient lensing configurations can produce multiple
images of the same background galaxy or quasar. Each of these images is further
distorted by intervening small-scale structures, thus the differences in their
shape and/or flux can be used to constrain the substructure content of galaxy
and cluster haloes
\citep{MaoSchneider1998,Metcalf2001,Dalal2002,Kochanek2004,Natarajan2004,Natarajan2007,Vegetti2010}. This
method is, in fact, the only way of detecting substructures in distant galaxies
\citep{Vegetti2012}. The amount and compactness of halo substructures depends
strongly on the nature of the DM particle: colder candidates produce more and
denser substructures
\citep[e.g.][]{Moore1999b,Klypin1999,Diemand2007,Springel2008}; particles with
larger self-interaction cross-sections produce shallower and more spherical
density profiles \citep{Meneghetti2001, Peter2013}. Therefore, strong
gravitational lensing can probe the microphysical properties of the DM particle
and thus provide a direct test of the Cold Dark Matter (CDM) paradigm.

In order to fully exploit gravitational lensing measurements in both strong and
weak regimes, it is essential to have accurate predictions for the nonlinear
state of the mass distribution in the Universe. In particular, for the
abundance, spatial distribution and internal properties of DM halos and their
substructure. Among the different theoretical approaches available,
cosmological $N$-body simulations appear as the only robust and accurate method
that meets these requirements.

Moreover, cosmological $N$-body simulations
\citep[e.g.][]{Peebles:1971a,Efstathiou1981, Efstathiou1985, Springel2005a,
Angulo2012a} are also invaluable cosmological tools: i) They are the most
reliable and precise method to follow the highly nonlinear evolution of
primordial density fluctuations \citep[e.g.][for a recent review]{Kuhlen2012a}.
ii) They provide virtual universes with which we can test, predict, and
interpret astronomical observations \citep[e.g.][]{Overzier2013}. iii) They
allow us to experiment with the laws of physics and the background cosmological
model \citep[e.g.][]{Fontanot2012,Fontanot2013}. Thus, numerical simulations
not only can provide the theoretical predictions required by gravitational
lensing, but also can be particularly useful for testing analysis algorithms
and for exploring the connection between lensing observations and the
underlying cosmological model \citep{Bartelmann:1998, Jain:2000, ValeWhite2003,
Meneghetti:2007, Hilbert:2009}.

Unfortunately, numerical simulations have a serious limitation that is inherent
to the formulation of the $N$-body problem: in order to efficiently solve the
Poisson-Vlasov equation, the initial cosmic density field must be represented
by a set of {\it discrete} bodies. This discretization is necessary to
efficiently follow the dynamics and evolution of the DM fluid, but it
introduces a small-scale noise that is very often larger than the small-scale lensing
signal itself. The noise decreases on large scales and/or with better mass
resolution. However, it is still comparable to the strong lensing signal from
most of the substructure population, even with the highest resolution
simulations to date \citep{Xu2009,Rau:2013}. In other words, the substructure
lensing properties that could allow us to constrain the DM particle mass remain
buried beneath the discreteness noise. Hence, current theoretical predictions are
not sufficiently accurate for upcoming lensing measurements.

In this paper, we propose \rtcm{}\footnote{The term \rtcm{} abbreviates for 
"Recursive deposit of Tethrahedra approximated by their Center of Mass".}, a
method to create gravitational lensing simulations free of discreteness noise,
with no tunable parameters nor additional smoothing scales. Our procedure
builds on a recently proposed method to solve for the collisionless dynamic of
the DM fluid \citep{Abel2012, Shandarin:2012, Hahn2013, Kaehler2012,
Angulo2013a}. The novel approach considers simulation particles as the vertices
of Lagrangian phase-space volume elements, not mass carriers as in the usual
interpretation of numerical simulations. The evolution and distortion of these
volume elements is described by the Eulerian coordinates of simulation
particles. Consequently, the DM density field is determined by the spatially
overlapping phase-space elements, these can be exactly deposited onto a target
grid using a recursive algorithm. The result is a continuous and smooth density
field ideal for small-scale lensing simulations.\footnote{The reduction of
discreteness noise also helps to suppress the artificial fragmentation of
filaments seen in warm Dark Matter (WDM) simulations
\citep{Hahn2013,Angulo2013a}.}

We devote this paper to the presentation and testing of the algorithm. We start
in \Sec{s:lensing} by describing how we compute the gravitational lensing
signal of a set of simulation particles. We then apply our method to a
cluster-size halo simulated in CDM and WDM cosmologies. These simulations are
described in \Sec{s:simulations}. In \Sec{s:results}, we compare our method
with standard approaches, and show how the noise in the surface density and
magnification maps is greatly reduced.  This allows us to explore the impact of
substructure on the strong lensing magnification fields for our CDM and WDM
haloes. We present our conclusions and a discussion of possible future work in
\Sec{s:conclusions}.

\section{Lensing Simulations}
\label{s:lensing}

We start by describing how the gravitational lensing signal of a set of
simulation particles is computed, including details of our method to
estimate the respective surface density maps.

\subsection{Gravitational lensing}
\label{s:sub_lensing}

Within the plane lens approximation, the lensing distortions produced by
a concentrated mass distribution can be derived from a lensing potential, $\Psi(\vtheta)$,
\citep[e.g.][]{BartelmannSchneider2001_WL_review}
\begin{equation}
\label{eq:lens}
\Psi(\vtheta) = \frac{1}{\pi} \int \diff{^2\vtheta}\kappa(\vtheta) \ln\left|\vtheta - \vtheta'\right|
,
\end{equation}
 where $\vtheta=(\theta_1,\theta_2)$ denotes an angular position on
the (plane) sky, and the convergence $\kappa(\vtheta)$ is defined as
\begin{equation}
\kappa(\vtheta) = \frac{\Sigma^\text{ang}(\vtheta)}{\Sigma^\text{ang}_\text{c}}
.
\end{equation}
Here, $\Sigma^\text{ang}(\vtheta)$ denotes the projected angular surface mass
density of the lens mass concentration. The critical angular surface mass
density is defined as
\begin{equation}
\Sigma^\text{ang}_\text{c} = \frac{\clight^2}{4 \pi \GNewton} \frac{a_\text{L} f_\text{L} f_\text{S}}{f_\text{LS}}
,
\end{equation}
with the speed of light $\clight$, gravitational constant $\GNewton$,
scale factor $a_\text{L}$ at lens redshift, and comoving angular diameter
distances $f_\text{L}$, $f_\text{S}$, and $f_\text{LS}$ from the observer to
the lens, from the observer to the source, and between the source and the lens,
respectively.

The deflection angle
$\vect{\alpha}(\vtheta)=\bigl(\alpha_1(\vtheta),\alpha_2(\vtheta)\bigr)$, the
complex shear $\gamma(\vtheta) = \gamma_1(\vtheta) + \ii \gamma_2(\vtheta)$,
and the magnification $\mu(\vtheta)$, are given by
\begin{align}
\vect{\alpha}(\vtheta) &= \bigl(\Psi_{1}(\vtheta),\Psi_{2}(\vtheta) \bigr) ,\\
\gamma(\vtheta)        &= \frac{1}{2} \left[\Psi_{22}(\vtheta) - \Psi_{11}(\vtheta) \right] - \ii \Psi_{12}(\vtheta) , \\
\mu(\vtheta) &= \left\{ \left[1- \kappa(\vtheta)\right]^2 - |\gamma(\vtheta)|^2\right\}^{-1}
.
\end{align}
where the subscripts refer to partial derivatives with respect to one
of the angular coordinates.

There are several ways of computing the lensing signal from numerical
simulations \citep[e.g][]{Wambsganss:1998,Couchman:1999,Jain:2000,Aubert:2007,
Hilbert:2009}. Here we choose one of the simplest, which consists in computing
the surface density on a regular lattice and then solving for the lensing
potential in Fourier space:
\begin{align}
\Psi^{F}(\vell)           &= \frac{1}{\pi} \kappa^F(\vell) \ln^F(\vell) \\
2\pi^2 \vell^2 \Psi^{F}(\vell) &= - \kappa^F(\vell)
\end{align}
where the superscript $F$ indicates a Fourier transform. These
expressions can be readily evaluated by using Fast Fourier Transforms (FFT).
However, this requires additional corrections, because FFT algorithms implicitly
assume periodic boundary conditions, while the appropriate conditions should be
vacuum boundary conditions. To suppress shear artefacts induced by periodic
images of the mass distribution, generous zero padding is employed.\footnote{For simplicity, we refrain from also applying a force-range cut-off.}
To recover the correct mean convergence [which is lost in the FFT
methods due to setting $\kappa^F(\vell=0)$ to zero], the potential from the FFT
is corrected by a term $\propto \vtheta^2$. Finally, the lensing deflection,
shear, and magnification can be obtained by computing derivatives of $\Psi$
either in Fourier space, or in real space by using finite difference methods \citep[][]{Hilbert:2009}.

\subsection{Recursive-TCM: A new density estimator}
\label{s:recursive_tcm}

The problem is now reduced to obtain the surface mass density,
$\Sigma(\mathbf{\theta})$, on a uniform grid from which the respective lensing
potential can be computed. Essentially, this step consists in mapping a
three-dimensional (3D) distribution of simulation particles onto a
two-dimensional (2D) grid. Although it is in principle a simple task, in
practice it is rather difficult to accurately carry out the mapping. Several
authors have explored different projection methods and have concluded that all
of them gave rise to a noise field of amplitude comparable to the strong
lensing signal produced by real DM substructures
\citep{Bradac2004,Li2006,Xu2009,Rau:2013}. This is true even for the highest
mass resolution simulations of DM haloes available to date. Similarly,
large-scale $N$-body simulations, with volumes comparable to that of future
wide-field galaxy surveys, have typically low number densities of simulation
particles, which adds a Poisson shot-noise that dominates the small-scale
predictions for weak gravitational lensing \citep[e.g.][]{Jain:2000,ValeWhite2003,Sato2009,Hilbert:2009}.

There are several proposed ways of dealing with this problem. For strong
lensing, one of the most common is to model the smooth component of a DM halo
with an analytic expression (e.g. a Single Isothermal Sphere), and then add on
top the substructure population \citep[e.g.][]{Xu2009}.  Although, it is
possible to incorporate the correct density profile and the triaxiality of the
DM halo, this method washes out all other higher-order or more subtle features
of the DM halo substructures such as streams, caustics, etc. For weak lensing,
maps are often smoothed with a fixed-size kernel, which decreases the
particle noise but also erases actual small-scale structure \citep{Hilbert:2009,Takahashi2011}.

Here, we discuss \rtcm{}, a mass-depositing
scheme that captures a simulated density field in all its complexity without
the noise introduced by the finite number of particles, and without any
smoothing parameter. The method extends the techniques proposed by
\cite{Abel2012,Hahn2013, Kaehler2012, Angulo2013a}, and thus we refer to them
for an extensive discussion of the method. The key
idea is to consider simulation particles as vertices of
Lagrangian phase-space tetrahedra. At any redshift, the particles indicate the
current positions of these vertices. To create surface density maps, the matter represented by these tetrahedra is distributed on a target mesh using a recursive splitting scheme.

One way of interpreting our method is that it assigns to each particle a
smoothing kernel whose {\it size and shape} are given by their Lagrangian (not
Eulerian as in most smoothing methods) neighbours. In particular, this kernel
is anisotropic and not even uniquely defined in an Eulerian space. We also note
that our method is conceptually different to those that project a Delaunay or
Voronoi tessellation built from the Eulerian particle distribution
\citep{Schaap2000,Bradac2004}.

The four main steps of \rtcm{} are:

\begin{itemize}

\item[] {\bf 1) Creating the Initial Tessellation}: First, we need to
define a set of disjoint Lagrangian phase-space elements that fully fill the
volume of a $N$-body simulation. In three dimensions, the most natural choice
is a Delaunay tessellation of the unperturbed particle distribution. The result
is a set of tetrahedra (six times more abundant than the number of particles)
whose corners are given by the simulation particles.\footnote{Constructing
the tessellation can be a computationally expensive task for state-of-the-art
simulations \citep[e.g.][]{Pandey2013}. However, this is trivial if the
particle distribution is arranged in a regular lattice (as opposite to a
glass-like distribution): each set of 8 grid points defines a cube that is
subdivided into six disjoint tetrahedra.} The connectivity of each tetrahedra
is fixed and stored (it can also be trivially recovered from the particles' ID
number in case the ID is related to the position of a particle in an
unperturbed lattice).

 \item[] {\bf 2) Reconstructing the Evolved Tessellation}: After the simulation has been
evolved and the particles moved to different locations, the initial set of
tetrahedra (which therefore also moved) is reconstructed using the stored
connectivity. The internal density of each tetrahedron is assumed to be
uniform, and the density field at any given location is simply given by all
those tetrahedra that intersect the target location. We note that it is also
possible to compute, at any point of space, other quantities besides the
density, such as the number of streams, the velocity dispersion tensor,
vorticity, etc. (Hahn et al in prep).

\item[] {\bf 3) Projecting the Density Field}: The next step is to compute the projected
density field on a grid, i.e, to map the tetrahedra onto a 2D regular mesh. The
simplest way, called \tcm{} by \citet{Hahn2013}, is to represent each tetrahedron by a single point
mass located at the center of mass. Another option is to represent each
tetrahedra by 4 particles, preserving the monopole and quadrupole of the parent
polyhedra \citep{Hahn2013}. Here we propose a more exact deposit, referred to
as {\rtcm}, which consists in recursively biparting each tetrahedron along its
longest edge. The process continues until all the child tetrahedra are
completely contained inside one grid cell, or a maximum number of levels in the
recursion is reached. Then, each child tetrahedra is subsequently represented
using a single particle of mass $2^{-l}m_{\rm tet}$ (where $l$ is the recursion
level and $m_{\rm tet}$ is the mass of the top tetrahedron) that is deposited
using a Nearest Grid Point (NGP) assignment scheme.

\item[] {\bf 4) Removing Density Biases}: Over the range of scales in which the mass resolution of a given simulation is
adequate to describe the evolution and distortion of Lagrangian phase-space
volumes, our method provides a very reliable proxy for the density field
\citep{Abel2012}. However, tetrahedron-based density calculations are biased if the distortion of an initial phase-space
volume can not be represented by linear transformations. This happens in two
situations. One is at the center of DM haloes, which have high densities and short dynamical times. As discussed in \cite{Hahn2013},  this has the net effect of densities being overestimated at the halo center, and underestimated at slightly outer regions. The second situation regards the tidal stripping of substructures, where some vertices of a given tetrahedron are stripped while
others might still be attached to the substructure. This has the
net effect of underestimating the mass associated to substructures.

Fortunately, these biases in the density are small and can be identified and
corrected for. Moreover, the centers of haloes are typically dominated by a
stellar component (specially in galaxy-sized DM haloes, where the observational
search for substructures is focused), and also are affected by baryonic
processes absent in DM-only $N$-body simulations (such as feedback, adiabatic
compression, etc).  Hence, any DM-only-based predictions need to be altered to account
for these and produce realistic lensing efficiencies \citep[e.g.][]{Xu2009}, so
an additional correction that remove biases of our density estimator can be
easily included.

Here we propose and use a simple way to remove the bias:
\begin{itemize}
\item[\bf{4.1)}] We first compute an unbiased 2D density map using a traditional (noisier)
estimator applied to the same object.
\item[\bf{4.2)}] We then apply a correction factor, defined as the average ratio between densities computed using a traditional estimator and the {\rtcm}, in bins of {\rtcm} densities. This aims to correct the overestimation of central densities.
\item[\bf{4.3)}] We apply an additional correction factor to account for spatially coherent, large-scale biases related to tidal stripping of substructures. This extra factor is set to the ratio of the density maps using the traditional and the \rtcm{} estimator (after the above correction is applied), both Gaussian smoothed to keep only large-scale modes.
\end{itemize}

As we will show in the next section (cf.\Sec{s:biases}),
this simple correction procedure eliminates most biases in the surface
density maps, preserving the reduced noise properties. We note, however, that
more sophisticated correction methods are possible and should decrease
the biases even further. Some possible extensions are applying corrections
to 3D densities instead of projected ones, and/or applying separate correction
factors for different substructures.

\end{itemize}

\section{Recursive-TCM in action}
\label{s:simulations}

For illustrative purposes, we now apply our new method to numerical simulations
of cold and warm DM cosmologies. We start by presenting these $N$-body
simulations, together with one particular DM halo on which we focus our
analysis. Then, we provide details of our density estimator when applied to
these simulated objects.

\subsection{Parent $N$-body runs}

We employ two of the cosmological $N$-body calculations presented in
\cite{Angulo2013a}, that simulate two different cosmological scenarios: i) a
standard CDM, and ii) a Warm-DM (WDM) model with a $250\,\eV$ DM particle mass.
In the latter, fluctuations below $k \sim 1\hMpc$ are suppressed, which
translates into a lack of collapsed structures below $M \sim
2\times10^{12}\Mass$, and consequently into a strong suppression of the subhalo
population of massive haloes. Although this WDM model is ruled out by
observations \citep{Viel2013}, we will consider it for illustrative purposes.
The cosmological parameters for the simulations are consistent with the WMAP7
data release \citep{Komatsu2011}: $\Omega_m = 0.276$, $\Omega_{\Lambda} =
0.724$, $\Omega_b = 0.045$, $h = 0.703$, $\sigma_8 = 0.811$ and spectral index
$n_s = 0.96$.

Each of these two simulations corresponds to a cubic region of $L=80\Mpc$ side
length, containing $1024^3$ simulation particles of mass
$3.65\times10^{7}\Mass$.  The initial conditions were created using the {\tt
MUSIC} code \citep{HahnAbel2011} at $z=63$. The particles were subsequently
evolved using a Tree-PM method, as implemented in the {\tt L-Gadget3} code
\citep{Angulo2012a,Springel2005a}.  Gravitational forces are smoothed using a
Plummer-equivalent softening length set to $5\kpc$. Additionally, we have
located DM haloes using a FoF algorithm \citep{Davis1985} (using a standard
value for the linking length $b=0.2$), and identified self-bound substructures
(or subhalos) within these haloes using the {\tt SUBFIND} algorithm
\citep{Springel2001a}.

The numerical simulations were started using identical phases and evolved with
the same numerical parameters, which allows a direct comparison of structure
formation in general, and of the gravitational lensing signatures explored in
our paper.

\subsection{Target cluster-sized DM halo}

For our strong gravitational lensing analysis, we will focus on the most massive
cluster present in our simulations at $z=0$. This object has a mass of $M_{200}
= 4.38\times10^{14}\,\Mass$, and it is resolved with more than $10$ million
particles. For comparison, this corresponds roughly to the lowest resolution
runs of the clusters in the Phoenix project \citep{Gao2012}, and it is a factor
of $10$ coarser than the main cluster employed by \cite{Rau:2013}.\footnote{Our force resolution is also much lower.}
The force resolution is $\sim250$ times smaller than the halo's virial radius, and thus the halo
structure is resolved adequately for our purposes.

The spherically averaged density profile of the halo is well fitted by a cored NFW profile
\citep{Navarro1996,Navarro1997}, $\rho(r)^{-1} \propto \sqrt{(r/r_s)^2  +
(r_c/r_s)^2} ( 1 + r/r_s)^2$, with concentration $r_s/r_{200} = 3.9$, and core
radius, $r_c = 5\kpc$, both in CDM and WDM. We note that the core is a
numerical artifact, and arises due to the lack of force resolution on scales
smaller than the simulation softening length.

\begin{figure*}
\centering
\includegraphics[width=0.44\linewidth]{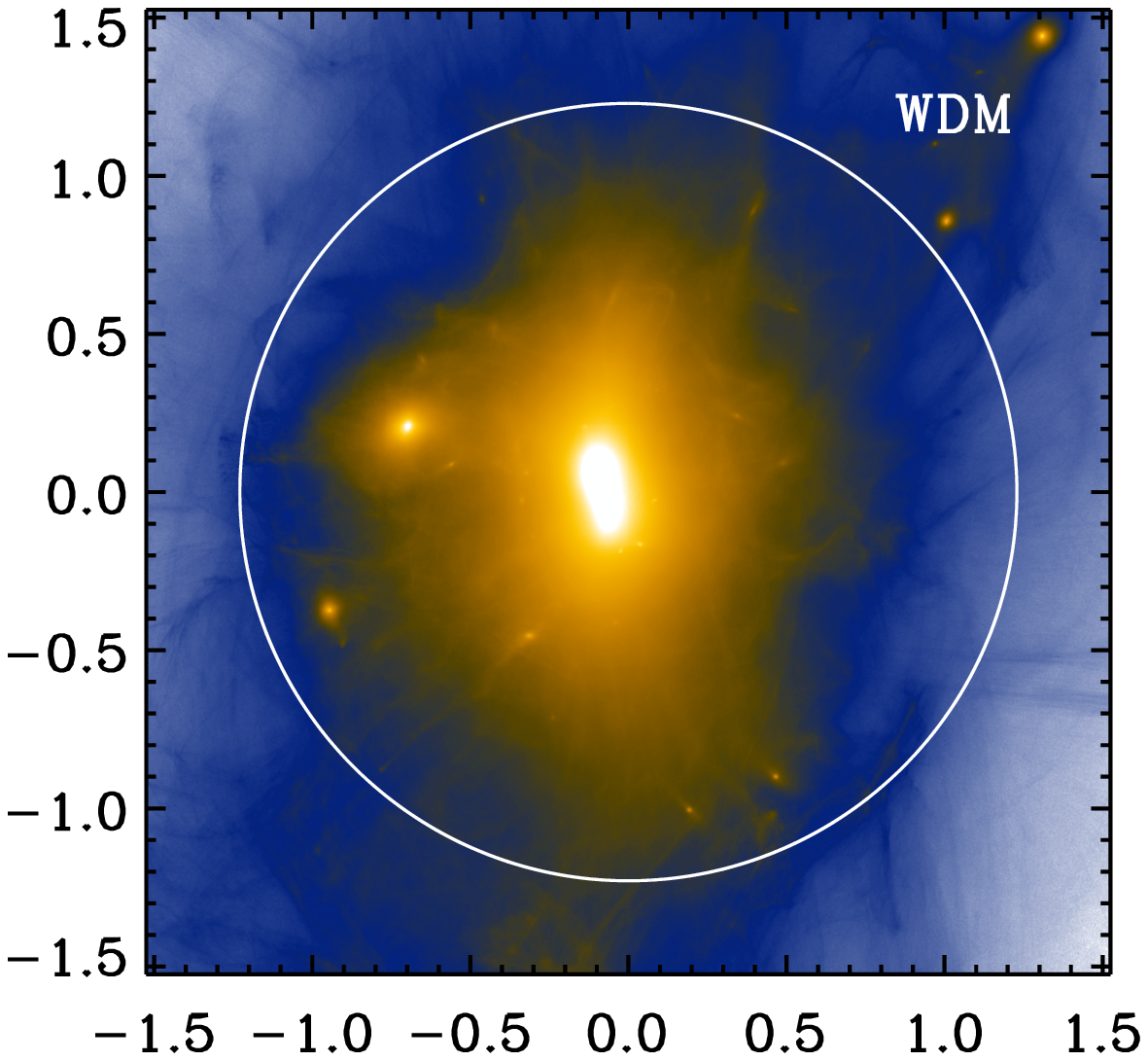}
\hfill
\includegraphics[width=0.44\linewidth]{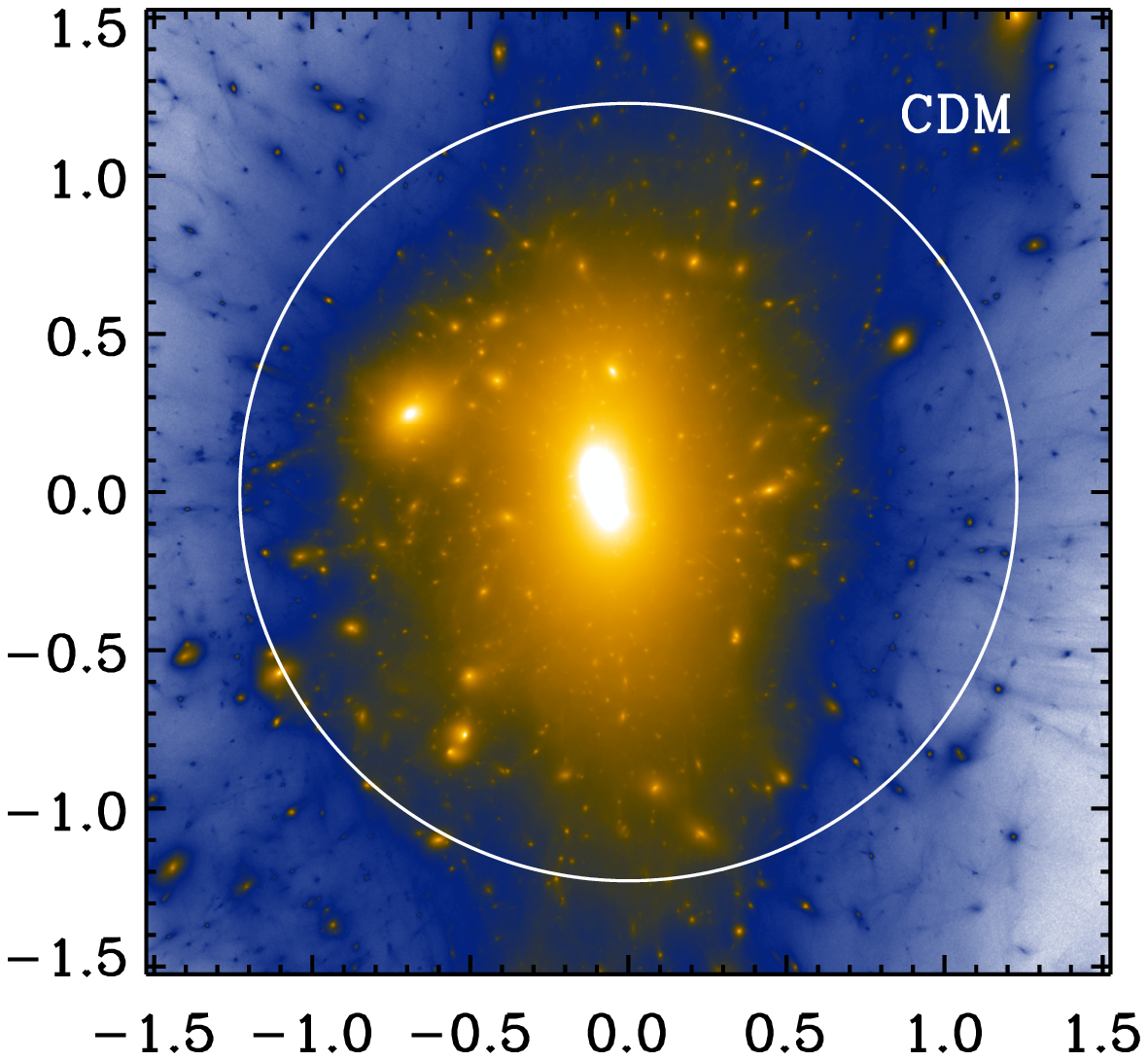}
\raisebox{0.12\height}{\includegraphics[width=0.1\linewidth]{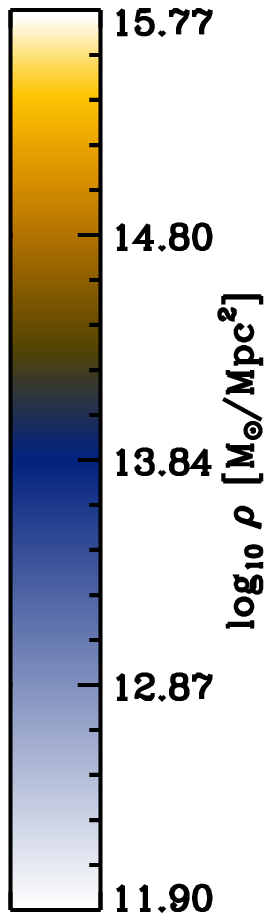}}
\caption{
Projected DM density, as computed by \rtcm{}, for the most massive halo in our
simulations at $z=0$ ($M_{200}=4.38\times10^{14}\Mass$).
Each image corresponds to a square region of $3.05\Mpc$ side length and $3\Mpc$ projection depth around the cluster center.
The white circle indicates the virial radius $R_{200}=1.229\,\Mpc$.
The left panel shows the halo in a WDM scenario, the right panel
assumes a CDM cosmology. \label{fig:halo}}
\end{figure*}

Fig.~\ref{fig:halo} shows an image of the selected halo in our two cosmological
scenarios. The DM halo displays the same overall structure in WDM and in CDM, and the
differences caused by the DM particle mass are evident only in their small-scale
properties. In CDM, the halo contains a wealth of substructure: a large number
of small clumps that are the remnants of previously accreted DM haloes.  These,
in contrast, are almost absent when WDM is adopted, but caustics, streams and
filaments are instead much better defined. Inside $R_{200}$ of the CDM halo,
we find $2121$ substructures with more than $15$ particles, which corresponds
to a minimum subhalo mass of $M_s \ge 7.3\times10^8\Mass$. Contrasting this, we
found only $119$ substructures inside the WDM halo -- which are mostly a result
of numerical fragmentation of filaments \citep{Wang2007a, Angulo2013a}. The
substructure population contributes $1.7\%$ and $6.5\%$ of the mass inside
$R_{200}$, respectively for our WDM and CDM halo.

Considering only substructure with masses above $10^{10}\,\Msun$, the subhalo
mass function in CDM follows a power law $dn/d\log{m_s} \propto m_s^{-0.79}$.
However, the slope decreases to $-0.66$ when we consider all the subhalos
detected. These values are shallower that the average slope found in other
simulations \citep[-0.9, e.g.][]{Angulo2008b, Gao2012}. The discrepancy is most
likely caused by our low force resolution compared to our mass resolution (many
low-mass haloes are tidally disrupted too efficiently due to our low force
resolution, which makes our subhalo mass function being incomplete up to higher
subhalo masses than in simulations with smaller softening lengths), which could
explain the change in slope in the subhalo mass function. Although these
discrepancies are not important for our work, we caution the reader that the
amount of substructure in our work is underestimated compared to other
simulations of similar mass resolution.

\subsection{Recursive-TCM lensing simulations}

We are now in position of applying our method to the WDM and CDM halo described
before. We artificially place the haloes at $z=0.32$, where the most massive
galaxy clusters are expected to be observed, and assume a background source
population located at $z=2$. We consider the 3D particle distribution inside a
region of dimensions $0.6\times0.6\,\Mpc$ (equal to $0.5 \times R_{200}$) and
$3\,\Mpc$ deep centred on our halo, and project it onto a $1000^2$-pixels
mesh.  This yields a spatial resolution of $0.6\kpc$, sufficient to resolve the
smallest structures present in our simulation given our gravitational
resolution limit ($5\kpc$).

We apply our full {\rtcm} algorithm using a maximum of ten recursion levels,
i.e. every top-level tetrahedron is split into $2^{10} = 1024$ smaller
tetrahedra, at most. Using these maps, we create convergence fields, compute
the lensing potential, and derive the associated $\mu$, $\gamma$, and $\alpha$,
as described in \Sec{s:sub_lensing}. We use a $32768^2$ FFT mesh (this is
approximately a factor of thousand larger than the density mesh to allow for
non-periodic boundary conditions), and compute the spatial derivatives in Fourier space. We note that by considering only a small inner region, we
are neglecting the weak lensing effects of structures further away from the cluster center.  However, we have explicitly
checked that this is a good approximation for the quantities explored here.

The computational cost of our \rtcm{} algorithm is higher than usual projection
methods, but it is still negligible compared to the time employed in carrying
out an $N$-body simulation. For our particular data structure, data access and
target grid, and maximum recursion level, it took $123$ minutes using $256$
processors, i.e. $500$ CPU hours. It is important to remember that this
corresponds to an implementation in software for CPUs, and that less recursion
levels significantly reduce the execution time. In addition, alternative
algorithms  based on GPU-rendering routines can, in principle, achieve
significantly better performances \citep{Kaehler2012}.

For comparison, we computed densities using two other techniques, in addition
to {\rtcm}. The first one, referred to as {\cic}, represents each particle by a
cube of uniform density and size equal to the cell size of the target grid.
This is the most-common projection method in cosmology \citep{Hockney1981}. The
second method, referred to as {\sph}, employs a spherically-symmetric
polynomial kernel \citep{Springel2005a} to project each particle onto our 2D
grid. The characteristic size of the kernel is given by the local density
about each particle, explicitly, by the 3D distance to the 32-th nearest
neighbour. This approach is the core of the Smoothed-Particle-Hydrodynamics
\citep[SPH,][]{Monaghan1992} numerical formalism.

\subsection{Bias correction}
\label{s:biases}

As discussed in \Sec{s:recursive_tcm}, {\rtcm} densities can
be biased in regions where heavy winding of the primordial
phase-space sheet occurs. Fortunately, we can use a noisy estimator to
correct for such biases, as we will show. In practice, we follow the
procedure described in \S2.2 and apply a two-step
bias correction.

\begin{figure}
\includegraphics[width=\linewidth]{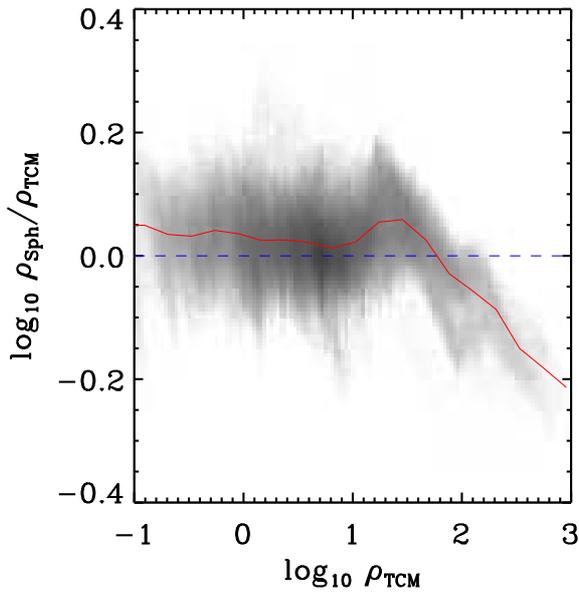}
\caption{Relation between the densities estimated using \rtcm{} and \sph{},
for $1000^2$ pixels covering the inner $0.6\times0.6\Mpc$ region centred in our
WDM cluster. The red line display the mean value in logarithmic bins of $\Delta
\log \rho_{\rm rtcm} = 0.1$.
\label{fig:corr} }
\end{figure}

The first step corrects for the overestimation of central densities by an average factor, shown in Fig.~\ref{fig:corr}. This factor was computed
as the geometric mean of the ratio between densities estimated using
{\sph} and {\rtcm}, in logarithmic
bins of $\Delta \log \rho_{\rm rtcm} = 0.1$. Because the densities produced
by our tetrahedral approach are biased high in central regions of DM haloes (see \Sec{s:recursive_tcm}), the average
ratio progressively decreases at higher densities. The largest correction
factor is $0.4$ at the very center of our halo.

\begin{figure}
\includegraphics[width=\linewidth]{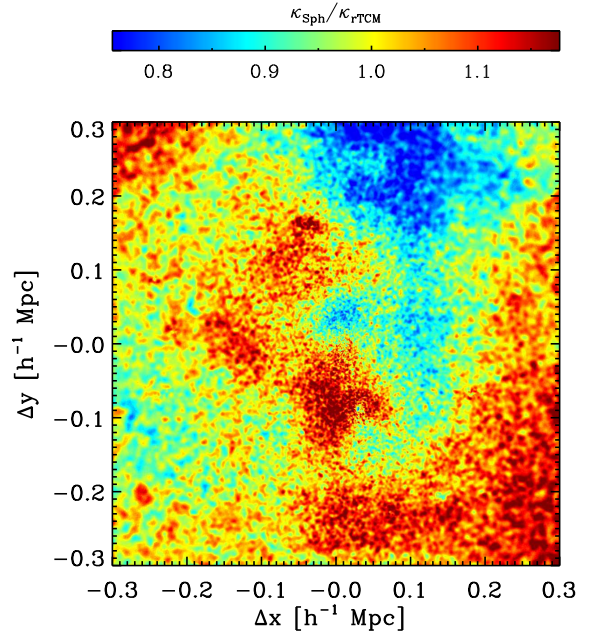}
\caption{
Fractional differences between the convergence maps built using the {\sph} and {\rtcm} methods. The region displayed correspond to the inner 
$0.6\times0.6\Mpc$ centered in our WDM halo.}
\label{fig:kappa_diff}
\end{figure}

The second correction step accounts for spatially coherent biases.
In Fig.~\ref{fig:kappa_diff}, we show the ratio between convergence maps
in {\sph} and in {\rtcm} for our WDM halo and after the first correction
has been applied. Ideally, this figure would be a pure white-noise field,
however, we can clearly see that there is a large-scale component in this
field. This arises partially because the 2D projected density field is 
not a one-to-one function of the full 6D phase-space structure (which determines the amount of winding and overestimation). Another aspect contributing
to this map is related to mass accretion history and shortcomings of the
tetrahedron-based densities to represent the tidal stripping of infalling
DM halos. 
In order to correct for this, we further divide the {\rtcm} map
by the a version of the map shown in Fig.~\ref{fig:kappa_diff}, but smoothed with a Gaussian kernel of size $50\kpc$. 
We note that this scale is set to be larger than those dominated by discreteness
noise in {\sph}. As we will see later, the simple procedure described
here, is successful in producing accurate lensing maps.

\section{Results}
\label{s:results}

We now present and discuss the results of our lensing simulations. We first
focus on the input surface density maps, and then on the lensing magnification.
In particular, we discuss the performance of our algorithm and the role of
discreteness noise compared to the signal of DM substructures.

\subsection{The surface densities and lensing convergence}

\begin{figure*}
\includegraphics[width=0.91\linewidth]{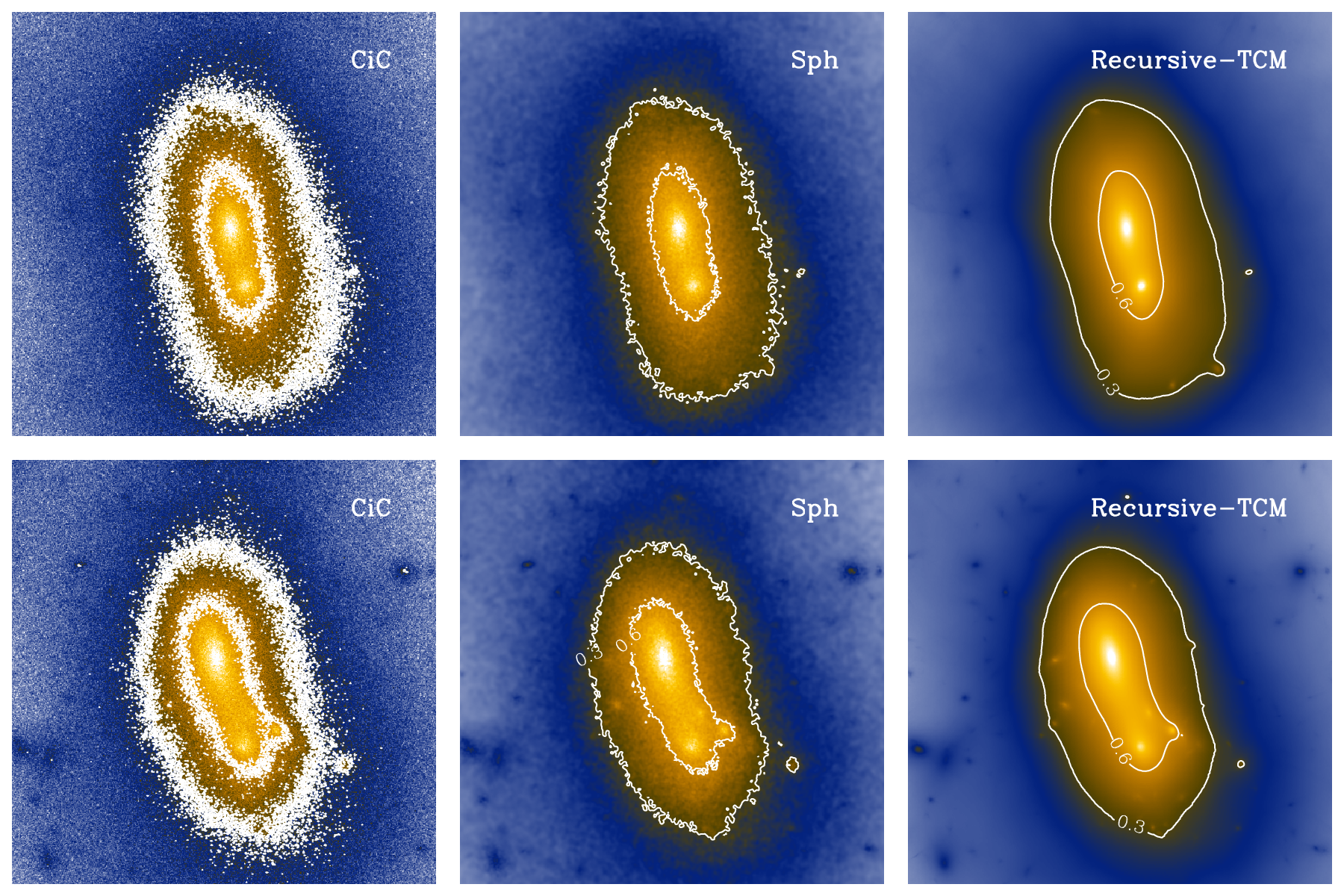}
\includegraphics[width=0.08\linewidth]{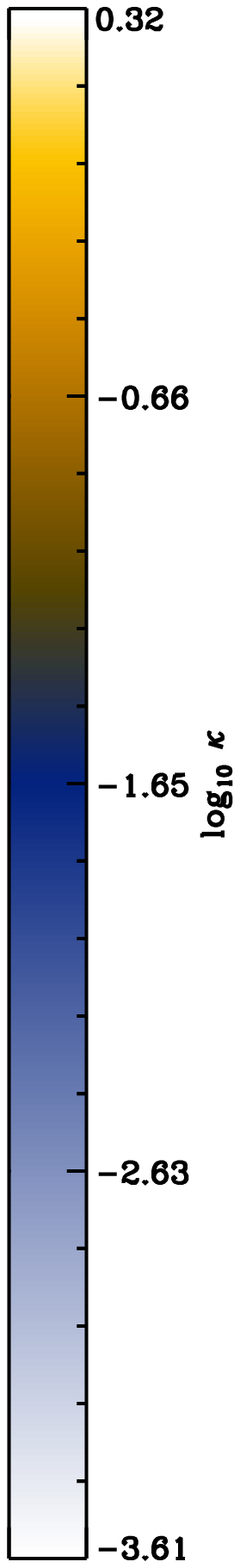}
\caption{The convergence $\kappa$ in the central $0.6\times0.6\Mpc$ region about our WDM
(top row) and CDM clusters (bottom row), using \cic{}, \sph{}, or \rtcm{} density estimates.
The colour scale is identical for all six sub-images. Note the different noise levels present in the different projection
methods. {\rtcm} displays the least noise.
\label{fig:inner_density}}
\end{figure*}

Fig.~\ref{fig:inner_density} shows maps of the surface density in the inner
regions of our WDM (top row) and CDM haloes (bottom row). Each column shows the
result of one of our three projection methods, as indicated by the legend.  Note
that the colour scale is identical in all six panels.

In both cosmological scenarios, we can appreciate how the small-scale noise is
decreased from left to right. The {\cic} method shows the largest fluctuations,
though we note that the visual impression of the noise depends on the target
mesh resolution, as this sets the size of the CiC smoothing kernel. The {\sph}
method reduces the noise significantly in this case, though still a considerable
amount of spurious fluctuations remains. These two images illustrate the
discretisation-related noise in traditional density estimators.

In contrast, the new {\rtcm} method, does not present any appreciable
noise thanks to the extra sophistication in the mass deposit, nor presents
appreciable biases thanks to the correction method described in
\Sec{s:recursive_tcm}. We note that despite being much smoother, all those
peaks seen clearly in the {\cic} and {\sph} maps also appear in the {\rtcm} maps.  
It is
important to note that our method is the only one that could in principle
distinguish random fluctuations in surface density maps from those produced by
halo substructures: compare the differences between the CDM and WDM halo in
rightmost column, with the differences seen in the leftmost column.  In both
SPH and CiC it is almost impossible to visually distinguish CDM from WDM.

\begin{figure}
\includegraphics[width=\linewidth]{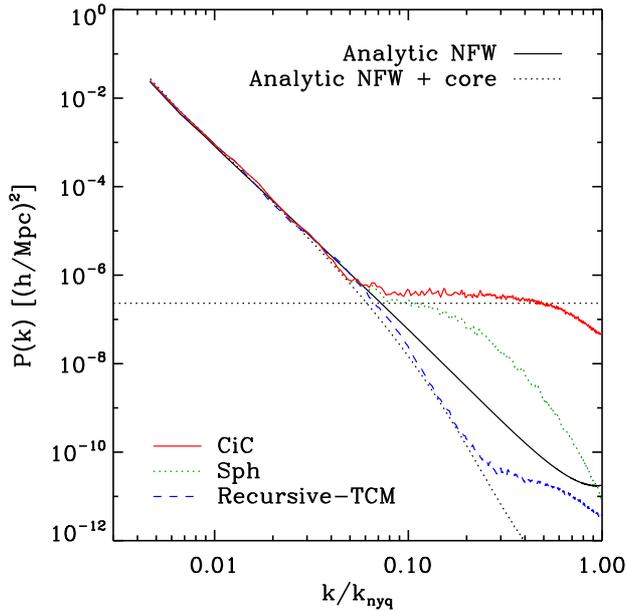}
\caption{The power spectra of the 2D over-density $\delta =
\Sigma/\langle \Sigma \rangle -1$, produced by the three
projection algorithms when applied to a WDM halo. Wavelengths are shown in units of the Nyquist frequency of the density mesh $k_{\rm nyq} = 5235\hMpc$. The method proposed here, {\rtcm}, yields the lowest amplitude on small-scale
fluctuations. Also shown are expectations for a smooth NFW profile without and with central core.
The dotted horizontal line indicates the white-noise level.
\label{fig:pk}}
\end{figure}

Fig.~\ref{fig:pk} shows the power spectrum of the surface over-density,
$\delta = \Sigma/\langle \Sigma \rangle - 1$, as given by the three
projection methods applied to the WDM halo.  We note that the measurements are
robust only until $k \sim 0.4 \times k_{\rm nyq}$,
where $k_{\rm nyq} = 1000 / (0.6 \Mpc) \times \pi = 5235\hMpc$ denotes the Nyquist frequency of the surface density mesh.
On smaller scales, aliasing
and the mass assignment window introduce visible artefacts, damping the
measured power spectra. We also display the expectations of a white-noise field
with the same number of point particles as those projected in our surface density
maps. For comparison, we also show the expectations for a (cored) NFW halo with
the same mass and concentration as the spherically-averaged density profile of
our DM haloes, but without noise.

Comparing all the measured power spectra, we observe a situation consistent with
the visual impression provided before. On large and intermediate scales, all
methods provide essentially identical power spectra decreasing as
$k^{-4}$, as expected for an NFW profile. This incidentally supports the validity
of our simple approach to correct the biases in {\rtcm} densities.

On scales smaller than $k \gtrsim 20\,\hMpc$ or roughly $r\sim50\,\kpc$ ($k >
0.08 \times k_{\rm nyq}$) -- a scale much larger than the typical size of
substructures -- all methods begin to differ.
The {\cic} spectrum follows the value expected for a 2D Poisson
field: $P_{\rm noise} = n^{-1} = 2.3\times10^{-7}$.
Interestingly, the {\sph} method levels to this expectation at roughly the same
scale as the {\cic} spectrum, but then decays much more quickly. This is because
the {\sph} method heavily smooths the field on scales smaller than the kernel
size. This smoothing also erases true
(specially anisotropic) small-scale density fluctuations present in our DM
haloes. For instance, it can be trivially seen that structures resolved with less
than $32$ particles will be smoothed out.

The performance of {\rtcm} appears to be considerably better. The noise level
is a few orders of magnitude below that of the other projection methods. The
noise measurements are consistent with our
expectations of reducing the noise in a way proportional to the  maximum level
of recursion, $l_{max}$, in the adaptive mass deposit: $P_{\rm noise} = n^{-1}
\times 6 \times 2^{l_{\rm max}}$, where $n$ is the number density of bodies
used in the map creation. The prefactor of six is understood in terms of the
six times more particles (one per tetrahedron) employed to describe the mass
field. This scaling predicts the Poisson noise in our measurements to be a
factor of $6 \times 2^{10} = 6144$ smaller than the {\cic} method. This value
is very close to the actual differences (measured at $k=0.27\hMpc$): $7136.7$.

Finally, we can see that {\rtcm} creates a projected mass power spectrum that is very close to that of an ideal NFW halo, differing only on the slope at high wavenumbers.
On these scales, the core introduced by the softening length in our
simulations becomes relevant, and the {\rtcm} power spectrum follows that of a cored NFW profile.

\subsection{Mass resolution dependence}

We now explore how the noise of our convergence maps vary with the mass
resolution of the underlying $N$-body simulation. For this, we have
down-sampled the field by factors of $2$, $4$ and $8$ in each coordinate, or
equivalently, reducing the total number of particles in our $N$-body
simulations by factors of $8$, $64$, and $512$. The most down-sampled case is
equivalent to a $128^3$ particle simulation, where our WDM halo would be
resolved with only $20000$ particles. For each case, we have created
convergence maps with maximum recursion levels set at $2$ for the original
maps, and to $5$, $8$ and $11$, respectively for the down-sampled versions. The
increased recursion level compensates the sparser particle data, so that in all
four cases the maps are created with roughly the same number of tetrahedra
(i.e. keeping $n^{-1} \times 2^{l_{\rm max}}$ constant).

\begin{figure}
\includegraphics[width=\linewidth]{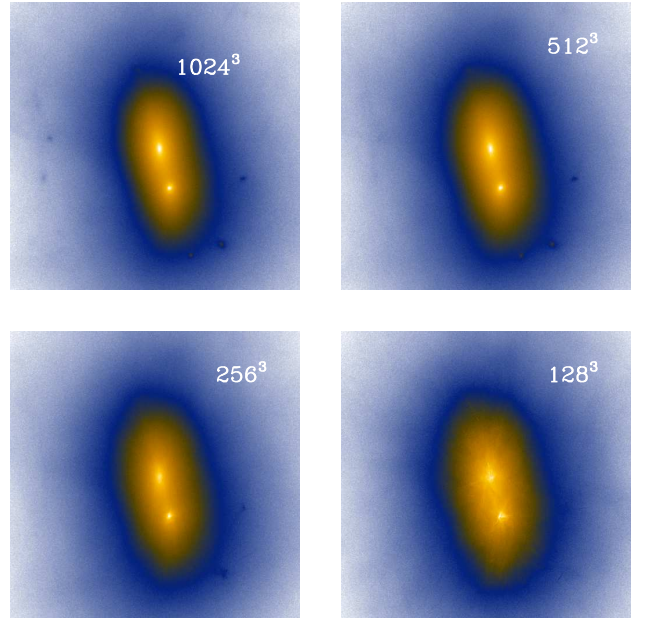}
\caption{ Comparison of the convergence maps created by the same particle
distribution of the WDM halo down-sampled by factors of $8$ (top-right), $64$ (bottom-left) or
$512$ (bottom-right). Therefore, these would correspond to haloes simulated with
$1024^3$, $512^3$, $256^3$ and $128^3$, respectively, as indicated by the
legend. The original map is shown on the top-left panel. We use an identical
logarithmic colour scale in all subpanels.
\label{fig:res_halos}}
\end{figure}

In Fig.~\ref{fig:res_halos} we show the four resulting convergence maps. In all
sub-panels we see that our method produces extremely smooth surface density
maps. Naturally, as we decrease the effective resolution, small-scale features
slowly disappear, for instance, the three substructures located on the
bottom-right side of the image. With low mass resolution, orbits and accretion
become discrete and subtle radial features appear. However,  even in the
lower-right case, which contains almost $3$ orders of magnitude less particles
than in our original simulation, the small-scale noise is considerably smaller
than with the {\cic} method (compare to the leftmost panel of the top row in
Fig.~\ref{fig:inner_density}). This shows that the limitation of our maps is in
the actual amount of structure that the parent $N$-body simulation tracks
correctly, and not in the discreteness noise associated with the finite number
of particles.

\begin{figure}
\includegraphics[width=\linewidth]{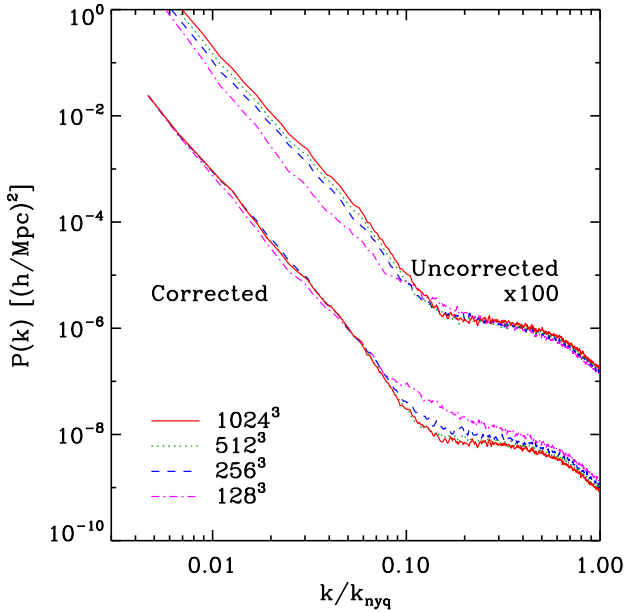}
\caption{ Comparison between the power spectra of projected overdensity maps
created from the WDM halo with the {\rtcm} method applied on progressively sparser data. The solid
red line show the result for our original dataset and a maximum level of
recursion set to $4$, whereas the dot-dashed magenta line shows the result for
a case using $512$ less particles but allowing seven further levels of
refinements.
\label{fig:res_pk}}
\end{figure}

\begin{figure*}
\includegraphics[width=0.91\linewidth]{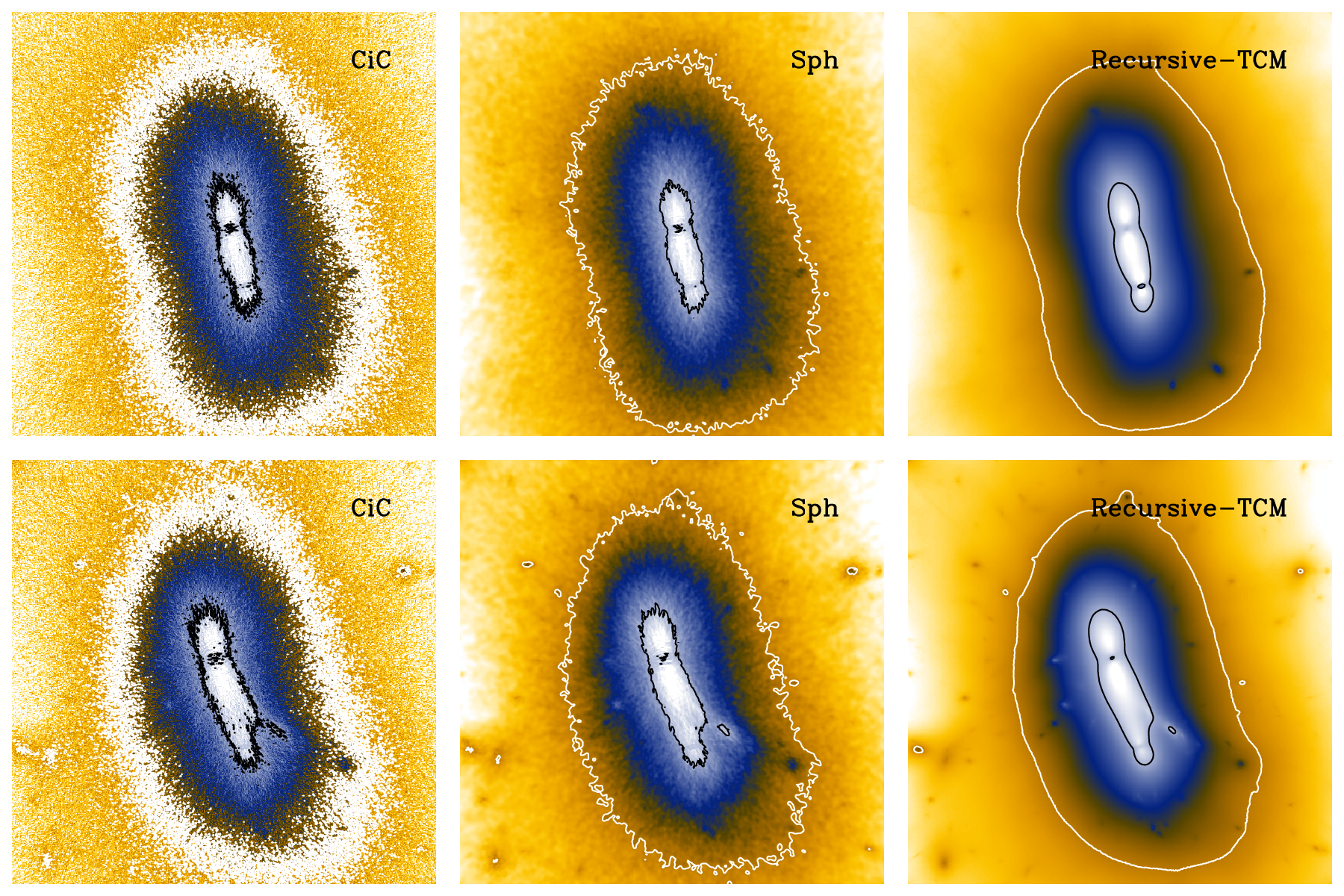}
\includegraphics[width=0.08\linewidth]{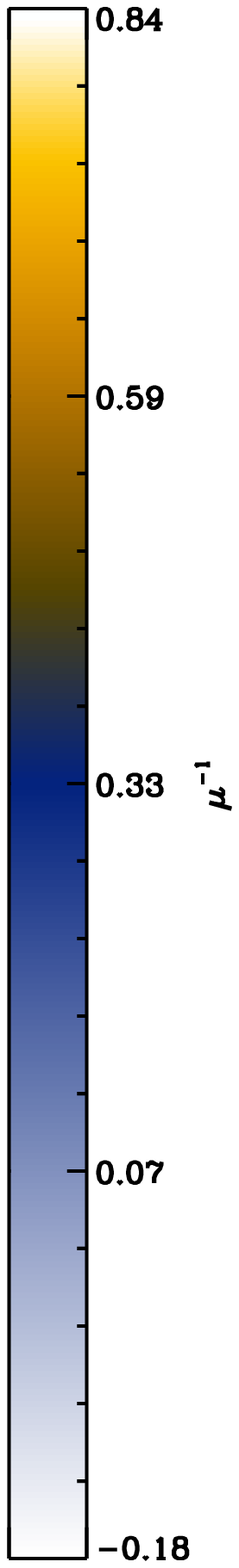}
\caption{ Map of the inverse of the magnification field, $\mu^{-1}$, at the
central region of our WDM (top) and CDM (bottom). The region displayed
matches that shown in Fig.~\ref{fig:inner_density}. White and black lines
shows contours where
$\mu^{-1}=0.6$ and $0$, respectively. Note we use the same colour scale
in all panels, ranging from $-0.18$ (white) to $0.85$ (light yellow).
\label{fig:imu}}
\end{figure*}

A quantitative comparison can be obtained from Fig.~\ref{fig:res_pk}, which
shows the 2D power spectra of our four test cases, with and without applying
our density correction (cf. \Sec{s:biases}). The overall shape of the power spectra is very similar
among all resolutions, as expected from the previous images, but small
differences arise due to the different amount of structure resolved in
the different cases. Nevertheless, the discreteness noise appears at the same
level and is set by the maximum amount of deposited tetrahedra. This again
shows that the limitation of our gravitational lensing simulations reside on
the ability of the parent $N$-body simulation to represent DM structures
properly, and not in an artificial noise introduced by our mass-projection
method. In the next subsection we will show that this has positive
repercussions on simulated lensing signals.

\subsection{The lensing magnification}

The magnification field, $\mu(\bf{x})$, which gives the ratio of the area of the
lensed image to the original area of the source, depends on second-order
derivatives of the lensing potential (whereas $\alpha(\bf{x})$ depends only on
first-order derivatives). Thus, $\mu(\bf{x})$ is very sensitive to small
perturbations to the lensing potential (such as those caused by subhalos). This
is also the reason why $\mu(\bf{x})$ is very susceptible to noise introduced by
discretisation in $N$-body simulations.

In Fig.~\ref{fig:imu} we display maps of the inverse of the magnification field,
$\mu^{-1}(\bf{x})$, created from each of the three projection methods we
consider. We highlight two places i) where the magnification is formally
infinite, $\mu^{-1}=0$, which are known as critical lines; and ii) where
$\mu^{-1} = 0.6$, which are useful to explore the impact of noise and
substructures in the topology of the magnification field. Note that our simulated
cluster is not a particularly efficient lens, partly due to its particular
dynamical state, and also because of our modest force resolution and the lack of
a modelling of a central stellar component.

\begin{figure}
\centerline{\includegraphics[width=\linewidth]{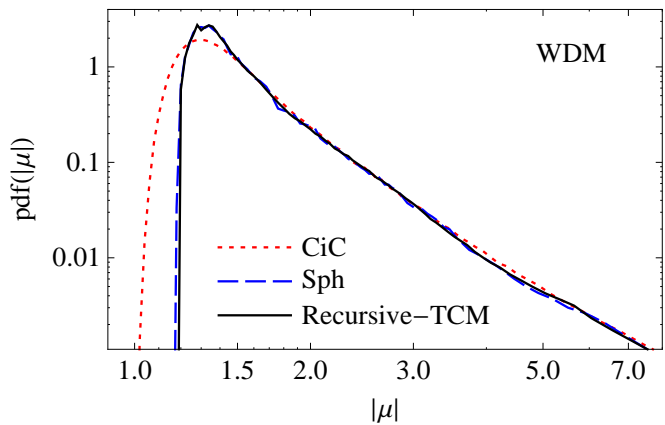}}
\caption{
Probability density $\pdf(|\mu|)$ of the magnification modulus $|\mu|$ computed
with {\cic} (red dotted), {\sph} (blue dashed), and {\rtcm} (black solid lines)
for the WDM halo.
\label{fig:mu_pdfs}}
\end{figure}

\begin{figure*}
\centerline{
\includegraphics[width=0.49\linewidth]{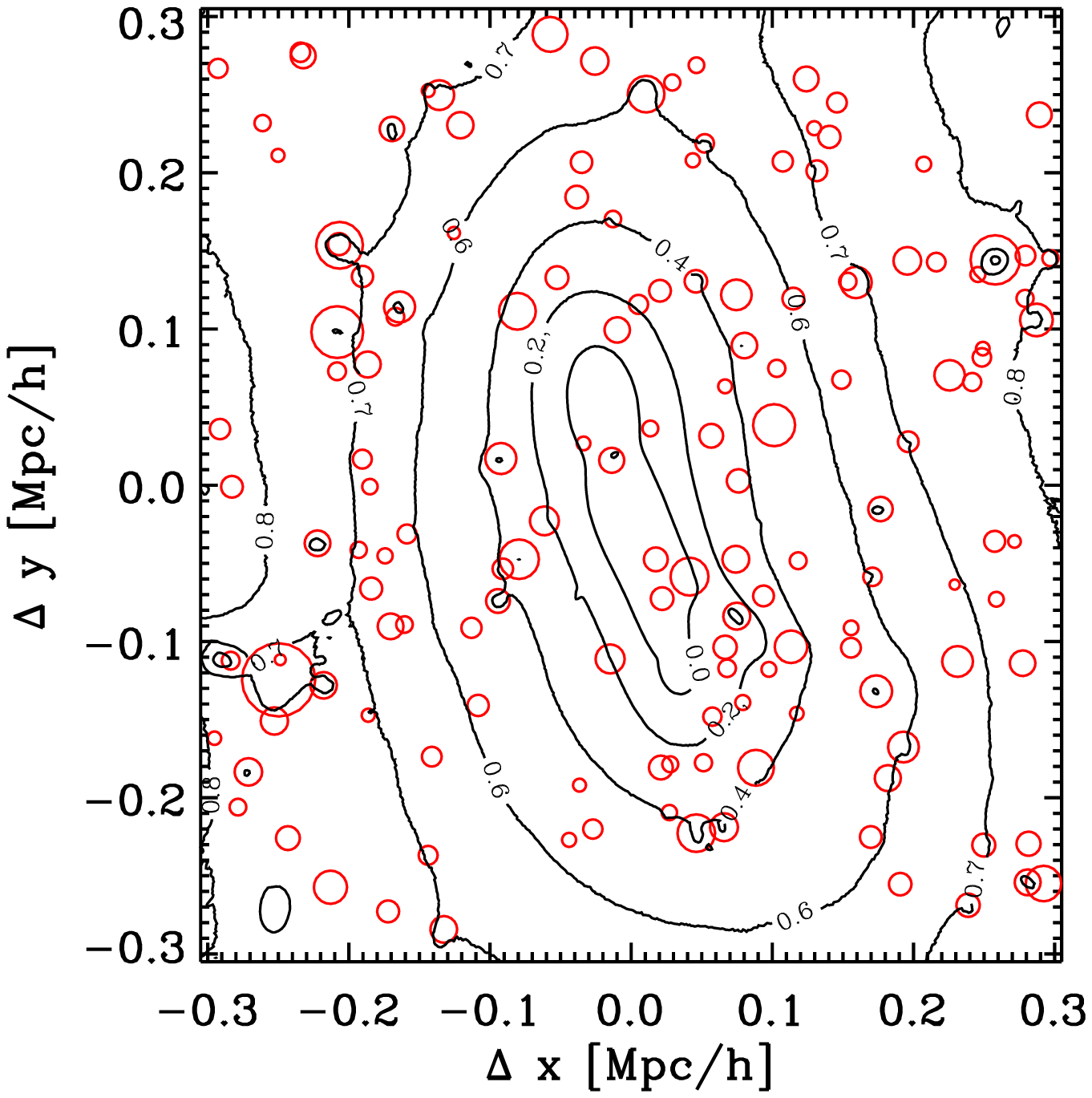}
\hfill
\includegraphics[width=0.49\linewidth]{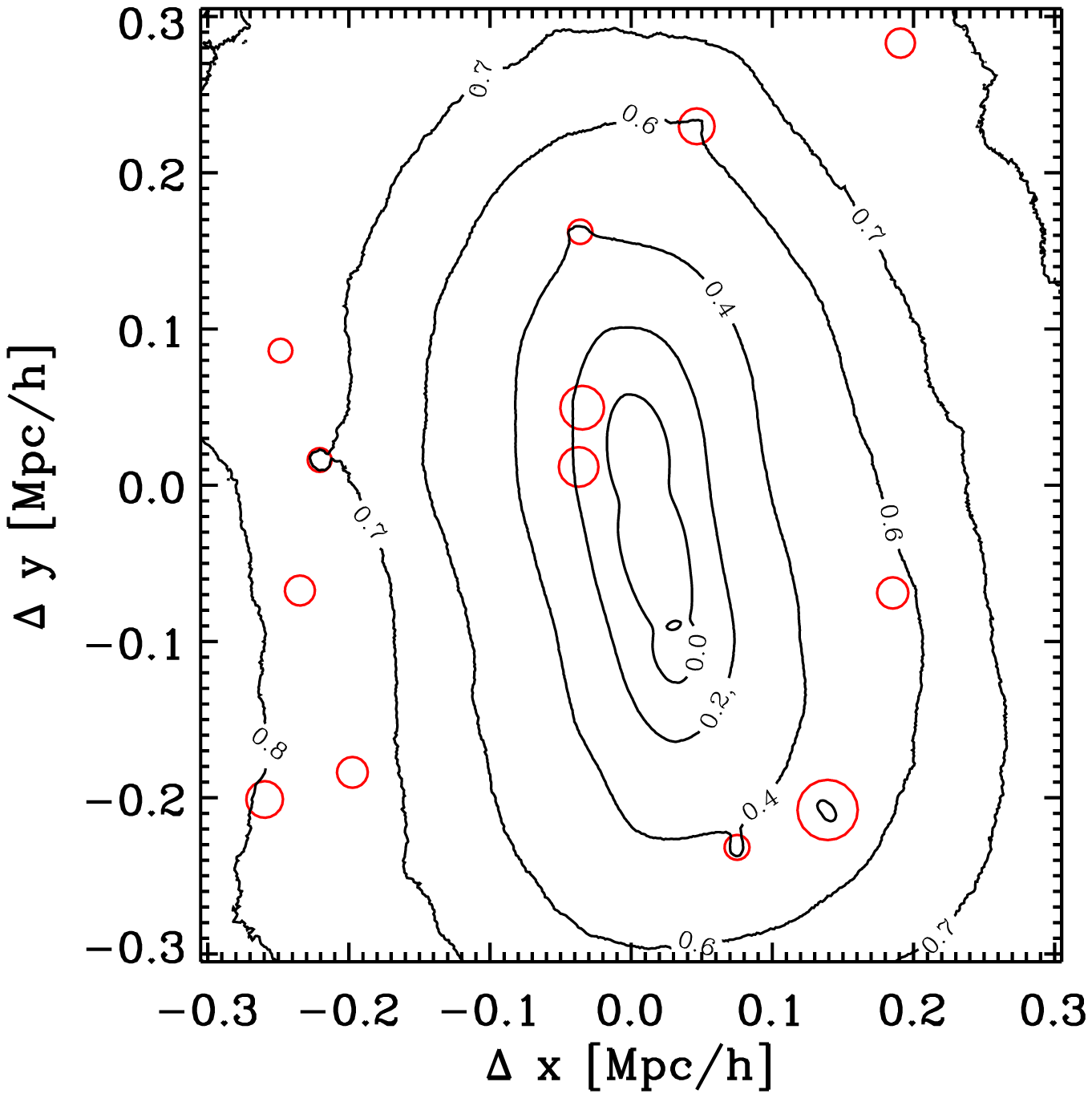}
}
\caption{The relation between substructure and perturbations in lensing
magnification for our simulated CDM (left) and WDM (right) halo. Black lines
denote iso-magnification contours at $\mu^{-1}=0.8,0.7,0.6,0.4,0.2$ and $0$
inwards. Red circles indicate the positions where substructures were
identified, and their radii is equal to the half-mass radius of the respective
subhalo. Note the reduced number of substructures in the WDM case, which result
from the initial suppression of small-scale fluctuations.
\label{fig:sub_pos}}
\end{figure*}

While the magnification maps from {\cic}, {\sph}, and {\rtcm} agree on large
scales, they differ substantially in the amount of associated small-scale
noise. In the {\cic} case, the noise fluctuations make it almost impossible to
distinguish CDM from WDM based solely on the topology of iso-$\mu$ lines. The
same is true to some degree in for {\sph}. The method {\rtcm} is superior: The
contours are not disturbed by discreteness noise. This allows us to explore the
magnification field with great detail. When applied to the WDM case, we see
contour lines that are extremely smooth. For the CDM halo, the contour lines
are also very smooth in most parts of the map, but display many small
protuberances with high significance. As we will see in the next section, these
are caused by the rich substructure population of this halo and thus are a
distinctive signature of the DM candidate properties.

Figure~\ref{fig:mu_pdfs} shows frequency of magnification values predicted by the different methods for the WDM (results for the CDM halo are very similar). The noise in the {\cic} magnification maps leads to substantially broader magnification distributions compared to those for {\sph} and {\rtcm}. In contrast, the distributions predicted by {\sph} and {\rtcm} are very similar. However, the probability densities predicted by {\sph} display more features, i.e. local deviations from a smooth density function, which we attribute to residual noise in the {\sph} maps.

\subsection{The impact of substructures}

In order to further explore the capabilities of our method, we now focus on the
impact of halo substructure on the magnification and convergence maps. In
Fig.~\ref{fig:sub_pos} we show iso-magnification lines, as
computed in {\rtcm} maps, together with the substructures
identified with more than $50$ particles in our simulated halo. 
Note that the small amount of noise visible in the outer contour is a result of
the maximum number of recursion levels ($l_{\rm max} = 10$) employed in our
method. This noise can also be seen in the power spectra of the projected
density field shown in Fig.~\ref{fig:pk}, and, as we discussed earlier, it can
be reduced further by simply increasing $l_{\rm max}$ at the expense of more
CPU time.

We can clearly see the differences between CDM and WDM. In the WDM case, there are only $13$ substructures in the
field, and consequently iso-$\mu$ lines are mostly smooth, showing only a few
notorious protuberances. In contrast, in the CDM case, there are $89$
substructures, and the iso-$\mu$ lines show many protuberances, but are almost
smooth otherwise.

As Fig.~\ref{fig:sub_pos} illustrates, all protuberances in the iso-magnification contours are associated with nearby
substructures. However, the relation is not simple, and different substructures
produce perturbations of different importance. Moreover, in some cases
fluctuations are not caused by a single substructure, but by a group of them.
This case is seen, for example, in the lower right section of the CDM
$\mu^{-1}=0.2$.  On the contrary, some substructures near contours do not
strongly perturb the magnification field. These objects have typically surface
mass densities below average.  For instance, the substructure located at
$(-0.04,0.05)$ in the WDM case, has a projected density a factor of ten smaller
than the substructure found at $(0.14,-0.2)$.

\begin{figure}
\includegraphics[width=\linewidth]{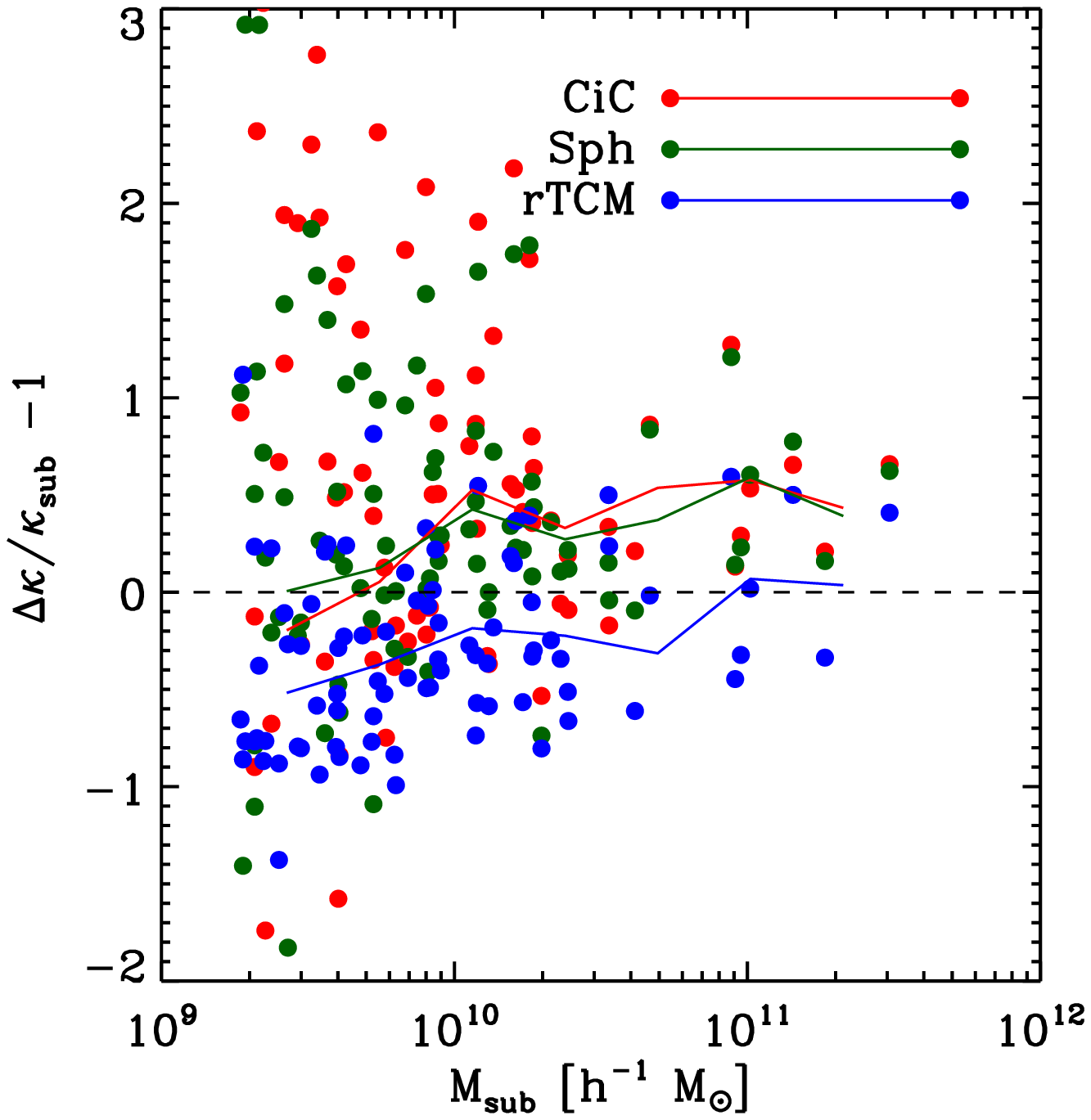}
\caption{
Ratio of the measured excess convergence $\Delta \kappa$ of substructures and a simple expectation $\kappa_{\rm sub}$, as a function of the subhalo
mass, $M_\text{sub}$, in maps constructed using the {\cic}, {\sph} and {\rtcm}.
Solid lines show the median values in seven logarithmic bins in
$M_\text{sub}$.}
\label{fig:subs_kappa}
\end{figure}

In Fig.~\ref{fig:subs_kappa} we compare the signal of identified
substructures among the convergence maps. The x-axis shows the true subhalo mass $M_\text{sub}$. The y-axis shows the
ratio of the excess convergence $\Delta \kappa$ with respect to a simple expectation $\kappa_\text{sub}$ based on
the substructure properties. The measured value, $\Delta\kappa$, is defined as $\kappa_\text{hm} - \kappa_\text{back}$, where $\kappa_\text{hm}$ is the mean convergence within the half-mass radius $R_\text{hm}$, and the background convergence $\kappa_\text{back}$ is the given by the mean convergence in an annulus with $1.5 R_\text{hm} < R < 1.7 R_\text{hm}$. 
The expected value, $\kappa_{\rm sub}$, is defined as $0.5 M_\text{sub}/(\pi R_\text{hm}^2)/\Sigma_c$.

The deviations of $\Delta \kappa/\kappa_\text{sub}$ from unity seen in  Fig.~\ref{fig:subs_kappa} can be explained, e.g., by particle noise (though not for \rtcm{}), projections effects, triaxiality, or inaccuracies in the background estimation.
Note that the scatter in $\Delta \kappa/\kappa_\text{sub}$ is much
lower in our method than in the other two. This
is thanks to a less noisy estimation of both the signal itself
and the background. However, the average value
of $\Delta\kappa$ is roughly a factor of two smaller in \rtcm{} than in the
other two methods. The discrepancy is originated by two factors. First, it
is due to an overestimation of the bias correction factor: 
since this factor is essentially set by the background halo, it does
not capture the exact density biases for substructures, which have
different central densities and dynamical times. The second aspect
is an intrinsic underestimation of the mass associated to
subhalos in {\rtcm}. This has an origin in tetrahedra being stretched
along a subhalo's orbit by tidal stripping. We estimated this effect to cause an
underestimation of about 30\% in the mass inside subhalos for resolved
substructures in our CDM halo. 
It remains to be explored whether more sophisticated correction
procedures, or modifications to the {\rtcm} algorithm, will alleviate these discrepancies.

\begin{figure}
\includegraphics[width=\linewidth]{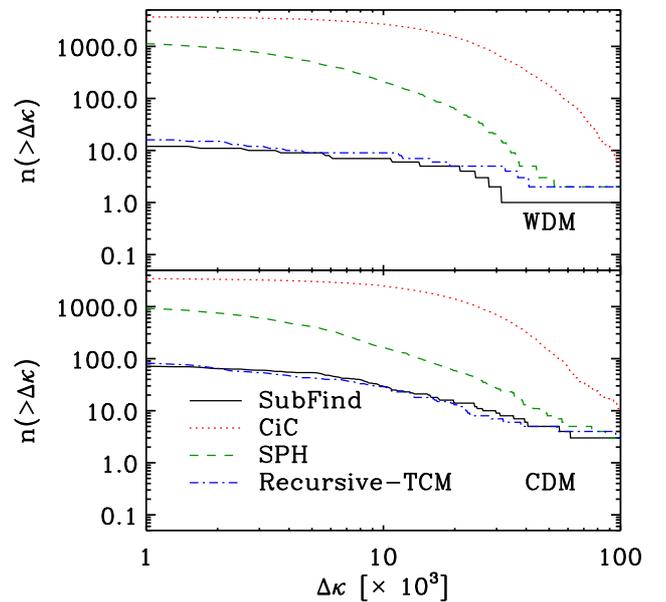}
\caption{ The cumulative number of local peaks in convergence maps detected
in our WDM (top) and CDM (bottom) cluster-sized halo. In each case we show the
results for a map created using three density projection methods:
{\cic} (red lines), {\sph} (green line) and {\rtcm} blue line. In
addition, we show the abundance of substructures detected in 3D by
the {\tt SUBFIND} algorithm.}
\label{fig:peaks}
\end{figure}

In order to quantify the performance of {\rtcm} concerning substructure
lensing signals, we have implemented a
hierarchical peak finder algorithm. This proceeds as follows: We start by smoothing the convergence field
with a Gaussian kernel of size $r_{\rm s} = 100\kpc$ and then identify local
peaks in the smoothed field. Then, we progressively reduce the kernel size and
search for new peaks, discarding those that are inside a larger peak. We repeat
this procedure for $20$ different scales, uniformly spaced in $\log{r_{\rm s}}$,
down to $r_s = 1\kpc$. Finally, we compute the signal associated to each peak as
$\Delta\kappa = \kappa_{\rm rs} - \kappa_{\rm back}$, where $\kappa_{\rm rs}$
is the average convergence within $r_{\rm s}$ and $\kappa_{\rm back}$ is the
local background value defined as the average of the map in an annulus of $1.5
 r_{\rm s} < r < 1.7 r_{\rm s}$.

The results are shown in Fig.~\ref{fig:peaks}, which displays the cumulative
number of peaks detected by our algorithm when applied to {\cic}, {\sph} and
{\rtcm} maps, as a function of their local convergence excess $\Delta\kappa$.
In addition, we plot as a black line the substructures detected in 3D by
{\tt SUBFIND}. The associated $\Delta\kappa$ is computed in the same way as
that of our peaks, but using $R_{hm}$ as the peak scale.
Our algorithm finds $4241$ (\cic), $1116$ (\sph) and $125$ (\rtcm) peaks in the
CDM map, and $4477$ (\cic), $1356$ (\sph) and $29$ (\rtcm), peaks in the WDM
map.

It is clear that the {\cic} and {\sph} maps contain a large amount of spurious
peaks produced by the discreteness noise. At all $\Delta\kappa$ there are
between one and two orders of magnitude more detected peaks than real
substructures. Moreover, the number of detected peaks is almost identical
between CDM and WDM, even though the actual amount of substructure is very
different. This further exemplifies that in current lensing simulations the
impact of noise is comparable or larger than that of real DM substructures.

In contrast, our method recovers roughly the correct amount of peaks,
which is an order of magnitude larger in CDM than in WDM.  Furthermore, $59\%$
and $69\%$ of the substructures can be matched to a peak in CDM and WDM,
respectively. Among those substructures not detected as peaks, we mostly find
objects with a negative or very small $\Delta\kappa$ value, which are also not
detected in the {\cic} or {\sph}. This suggests that these might indeed be the
false-positives in {\tt SUBFIND} or special cases of projection effects.
However, it might also be a consequence of the simplicity of our peak detection
algorithm. In order to quantify the implications for
observational constraints, this issue requires further study considering
realistic input signals and analysis procedures,
and perhaps more sophisticated procedures to correct for {\rtcm} biases.

\begin{figure}
\includegraphics[width=\linewidth]{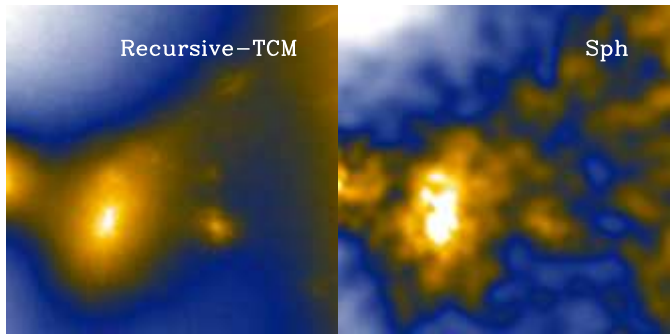}
\caption{ Zoom into a $100\times100\kpc$ region, centred at $(\Delta x,\Delta
y) = (-0.23,-0.11)\Mpc$ relative to the main halo, showing the convergence
field associated to a tidal feature. The left panel shows a map computed
using a {\rtcm} projection method, whereas the right panels shows one computed
using the {\sph} method.}
\label{fig:stream}
\end{figure}

There are also peaks in the convergence maps that are not related to any identified 3D
substructure. In many cases, these are due our 3D substructure finder
algorithm: a density peak not found by {\tt SUBFIND}, one that fell below our
mass-resolution, or one that is not a self-bound object. For instance, the large
peak located at $(-0.27,-0.28)$.
An object of a different nature is shown in
Fig.~\ref{fig:stream}. It does not correspond to a self-bound spherical
overdensity, but to a DM stream of a tidally
disrupted substructure. Such features are expected in hierarchical structure
formation scenarios, and perhaps they could be eventually detected trough their 
lensing
signal. (Note that this feature is barely distinguished over the noise in
{\sph}).  For the moment, this detection serves as a further example of the
potential accuracy and precision of the lensing maps created by the method
presented and discussed here.

\section{Conclusions}
\label{s:conclusions}

The next generation of gravitational lensing observations might help us to
decipher the mysteries of the Dark Universe: Dark Energy and Dark Matter.
However, high-precision theoretical predictions are essential to ensure the
correct interpretation of future datasets.

We presented {\rtcm}, a method aimed at predicting the lensing signal from
cosmological simulations with the required accuracy. This algorithm originates
from a novel way of interpreting the results of $N$-body simulations and
overcomes one of the most serious limitations of current lensing simulations:
the noise introduced by the discrete nature of the particle representation of
the density field.

We applied {\rtcm} to cluster sized-haloes simulated in WDM and CDM universes.
We showed that the method produces convergence maps with noise several orders
of magnitude below that of traditional methods (here, a factor of $\sim7000$
smaller than the formal shot-noise limit), and that it recovers the underlying
power spectrum of fluctuations well below the particle shot noise of the
simulations.

With traditional projection methods, the discreteness noise in lensing maps are
comparable to the signal from DM substructure. This is not true for {\rtcm},
where the features associated with real overdensities are preserved, but the
discreteness noise is absent.  All this comes without free parameters or
additional smoothing scales.  We also showed that there are density biases
associated to {\rtcm}, which, however, can be mostly eliminated by simple
correction procedures.  Therefore, this method is well suited for creating
high-precision predictions for the relation between the underlying cosmological
model and the expected signatures of small-cale structure in strong
gravitational lensing observations.

With {\rtcm}, we were able to clearly show the differences in the lensing
properties between CDM and WDM. The differences come mainly from their
substructure population, and thus lensing might be able to constrain the DM
particle mass. However, we found that the relation between substructures and
the associated lensing effects is not trivial: some substructures do not affect
the convergence noticeably; many lensing perturbations are caused by more than
a single structure; and a few perturbations are not associated with any
self-bound substructure, but, e.g. with tidal debris. This suggests that in
order to interpret correctly the lensing measurements of substructures, a
rigorous study of the detectability of substructures needs to be carried out.

In this paper, we have shown the feasibility of our method, providing examples
of its potential when applied to a rather modest simulation. In the future, we
expect our method to enable many detailed theoretical studies, exploiting
state-of-the-art simulations of much higher force and mass resolution, also
simulating more realistic WDM scenarios, and even taking advantage of
hydrodynamical simulations. The presented method will also be very useful for
creating large-scale weak lensing shear and magnification maps with high
fidelity and low particle noise. Moreover, the method will be crucial for
testing and characterising the performance of algorithms of extracting
substructure information from observed lensed galaxies, in particular methods
for constraining the DM particle mass from image perturbations or flux-ratio
anomalies in multiple-image systems. All this together will allow us to
understand better the impact of the underlying cosmological model in lensing
observations, and therefore help to unleash the full potential of gravitational
lensing.

\section*{Acknowledgements}
We warmly thank Oliver Hahn and Ralf Kaehler for many enlightening discussions.
We also thank Simon White for valuable suggestions that improved our paper.  We
acknowledge useful comments on the manuscript from Carlo Giocoli, Yashar
Hezaveh, Phil Marshal, Ben Metcalf, Stefan Rau and Simona Vegetti.  T.A.
gratefully acknowledges support by the National Science foundation through
award number AST-0808398 and the LDRD program at the SLAC National Accelerator
Laboratory as well as the Terman Fellowship at Stanford University.  We
gratefully acknowledge the support of Stuart Marshall and the SLAC
computational team, as well as the computational resources at SLAC. Parts of
the computations were carried out at the Rechenzentrum Garching (RZG) of the
Max Planck Society.

\bibliographystyle{mn2e} \bibliography{database}

\label{lastpage} \end{document}